\documentclass{article}
\usepackage{bm}
\newcommand{\vecb}[1]{\bm{#1}}
\usepackage{float}
\usepackage{amsthm}

\usepackage{PRIMEarxiv}

\usepackage[utf8]{inputenc} 
\usepackage[T1]{fontenc}    
\usepackage{hyperref}       
\usepackage{url}            
\usepackage{booktabs}       
\usepackage{amsfonts}       
\newcommand{\ii}{\mathrm{i}} 

\usepackage{nicefrac}       
\usepackage{microtype}      
\usepackage{lipsum}
\usepackage{fancyhdr}       
\usepackage{graphicx}       
\graphicspath{{figures/}}     

\theoremstyle{plain}
\newtheorem{theorem}{Theorem}[section]

\newtheorem{proposition}[theorem]{Proposition}

\theoremstyle{definition}

\theoremstyle{remark}

\usepackage{amsmath}
\usepackage{amsfonts}
\usepackage{amssymb}
\usepackage{amsthm}
\usepackage{graphicx,ifthen,calc}
\title{Complex-Shifted Regularized Regression for Ill-Conditioned and Rank-Deficient Linear Models}
\author{
  L.~H. Kuo and Achraf Cohen \\
  Department of Mathematics and Statistics\\
University of West Florida\\
Pensacola, FL 32514, U.S.A. \\
  \texttt{\{lkuo, acohen\}@uwf.edu} \\
}

\begin{document}
\maketitle

\begin{abstract}
Ill-conditioning and rank deficiency can produce unstable or nonunique regression coefficient estimates in linear models. We introduce Complex-Shifted Regularized Regression (CSRR), which replaces the real-valued perturbation of the least-squares normal equations by a purely imaginary shift. The resulting complex-valued system is nonsingular for every positive regularization parameter, including when the cross-product matrix is singular. Explicit expressions are derived for the real and imaginary components of the CSRR solution, and the real component is shown to admit a Tikhonov-type regularization interpretation. For full-rank designs, the real component converges to the ordinary least-squares solution as the regularization parameter approaches zero. For rank-deficient designs, it converges to the Moore--Penrose minimum-norm least-squares solution. For a fixed nonsingular problem and under the stated parameterization, the local perturbation of the CSRR real component is of order $O(\mu^2)$, compared with $O(\mu)$ for classical ridge regression with shift $\mu I_p$. Numerical examples involving polynomial and multiple regression illustrate these theoretical properties under well-conditioned, increasingly ill-conditioned, and exactly rank-deficient designs. The results also demonstrate the effects of finite-precision computation as ill-conditioning becomes more severe.

\end{abstract}

\section{Introduction}
\label{sec:introduction}

Linear regression is a foundational method for modeling the relationship between a response variable and a set of predictor variables \cite{MontgomeryPeckVining2012}. Throughout, we consider the linear regression model
\begin{equation}
\vecb{y}
=
X\vecb{\beta}
+
\vecb{\varepsilon},
\label{eq:linear-model}
\end{equation}
where $X\in\mathbb{R}^{n\times p}$ is the predictor matrix, $\vecb{y}\in\mathbb{R}^{n}$ is the response vector, $\vecb{\beta}\in\mathbb{R}^{p}$ is the vector of regression coefficients, and $\vecb{\varepsilon}\in\mathbb{R}^{n}$ is the random error vector. The predictors and response are assumed to have been centered, so that the intercept is treated separately and is not subject to regularization. The regression coefficients are obtained from the normal equations
\begin{equation}
X^{\top}X\vecb{\beta}
=
X^{\top}\vecb{y}.
\label{eq:normal-equations}
\end{equation}

When the columns of $X$ are linearly independent and $X^{\top}X$ is well conditioned, the least-squares solution can be computed reliably. Difficulties arise when the predictors are strongly correlated or when the design matrix is rank deficient. In such cases, $X^{\top}X$ may become severely ill conditioned or singular, resulting in unstable or nonunique regression coefficient estimates. Multicollinearity and related diagnostic issues are treated extensively in regression analysis \cite{BelsleyKuhWelsch1980,OBrien2007}, while numerical methods for ill-conditioned and rank-deficient least-squares problems are developed in \cite{Bjorck1996,GolubVanLoan2013,Hansen1998}.

A classical approach to reducing instability caused by multicollinearity is ridge regression, introduced by Hoerl and Kennard \cite{HoerlKennard1970,HoerlKennard1970Applications}; see also the historical review by \cite{Hoerl2020}. Ridge regression modifies the normal equations by introducing a positive regularization parameter,
\begin{equation}
\left(
X^{\top}X+\mu I_p
\right)
\vecb{\beta}_{R,\mu}
=
X^{\top}\vecb{y},
\qquad
\mu>0,
\label{eq:ridge-system}
\end{equation}
where $I_p$ is the $p\times p$ identity matrix, $\mu$ is the ridge regularization parameter, and $\vecb{\beta}_{R,\mu}$ is the ridge regression coefficient vector. Because $X^{\top}X$ is positive semidefinite, $X^{\top}X+\mu I_p$ is positive definite for every $\mu>0$. Ridge regularization therefore yields a uniquely defined regression coefficient vector while reducing sensitivity to ill-conditioning. Generalized ridge estimators, tuning-parameter procedures, and numerical investigations of ridge regression have subsequently been developed \cite{Marquardt1970,Hemmerle1975,HoerlKennardBaldwin1975,LawlessWang1976,GolubHeathWahba1979}.

Other regularization methods include the LASSO, which uses an $\ell_1$ penalty to combine coefficient shrinkage with variable selection \cite{Tibshirani1996}, while the elastic net combines $\ell_1$ and $\ell_2$ penalties and is particularly useful for correlated predictors \cite{ZouHastie2005}. Related developments include the nonnegative garrote, adaptive LASSO, and efficient algorithms for regularization paths \cite{Breiman1995,Zou2006,FriedmanHastieTibshirani2010}. These methods address objectives distinct from the complex perturbation studied here.

Complementary numerical approaches include singular value decomposition and generalized inverse methods for ill-conditioned and rank-deficient least-squares problems \cite{Moore1920,Penrose1955,Bjorck1996,GolubVanLoan2013,TrefethenBau1997}. Tikhonov regularization provides a general framework for stabilizing ill-conditioned and ill-posed problems, with classical ridge regression corresponding to zeroth-order Tikhonov regularization with an identity penalty matrix \cite{Tikhonov1963,Gockenbach2016}. These established approaches provide reliable solutions for a wide range of regression and inverse problems; the purpose of the present work is not to replace them, but to investigate a different perturbation of the least-squares normal equations.

Complex perturbations have recently been investigated as a regularization mechanism in function approximation. Kuo, Chen, and Karageorghis introduced a complex regularization method for radial basis function interpolation in which a purely imaginary perturbation is added to a singular or ill-conditioned real-valued system \cite{KuoChenKarageorghis2027}. Their results demonstrate that such a complex shift can regularize rank-deficient interpolation systems. Motivated by this construction, the present work extends the complex-shift approach to linear regression and studies the mathematical and numerical behavior of the resulting complex-valued regression solution.

We introduce \emph{Complex-Shifted Regularized Regression} (CSRR), defined by
\begin{equation}
\left(
X^{\top}X+\ii\mu I_p
\right)
\widetilde{\vecb{\beta}}_{\mu}
=
X^{\top}\vecb{y},
\qquad
\mu>0,
\label{eq:csrr-introduction}
\end{equation}
where $\ii=\sqrt{-1}$ and $\widetilde{\vecb{\beta}}_{\mu}\in\mathbb{C}^{p}$ is the complex-valued CSRR coefficient vector. We write
\begin{equation}
\widetilde{\vecb{\beta}}_{\mu}
=
\vecb{\alpha}_{\mu}
+
\ii\vecb{\gamma}_{\mu},
\label{eq:csrr-components-introduction}
\end{equation}
where $\vecb{\alpha}_{\mu}\in\mathbb{R}^{p}$ and $\vecb{\gamma}_{\mu}\in\mathbb{R}^{p}$ denote the real and imaginary components, respectively. Since $X^{\top}X$ is real symmetric and positive semidefinite, $X^{\top}X+\ii\mu I_p$ is nonsingular for every $\mu>0$, including when $X^{\top}X$ itself is singular. The real component $\vecb{\alpha}_{\mu}$ is used as the regularized regression coefficient vector, while the imaginary component $\vecb{\gamma}_{\mu}$ provides an accompanying path generated by the complex perturbation.

The present study has three objectives. First, we establish the fundamental mathematical properties of the CSRR system, including nonsingularity for every $\mu>0$, explicit representations of its real and imaginary components, and a real-valued Tikhonov interpretation of the real component. Second, we characterize the limiting behavior of CSRR as $\mu\rightarrow0$. For full-rank designs, the real component converges to the ordinary least-squares solution; for rank-deficient designs, it converges to the Moore--Penrose minimum-norm least-squares solution. Third, we use controlled numerical examples to examine these theoretical properties under well-conditioned, increasingly ill-conditioned, and exactly rank-deficient regression designs.

For a fixed full-rank regression problem and under the same stated parameterization, the perturbation of the CSRR real component from the least-squares solution is $O(\mu^2)$ as $\mu\rightarrow0$, whereas classical ridge regression with shift $\mu I_p$ has an $O(\mu)$ perturbation. This comparison describes the local dependence of the two estimators on their stated regularization parameters. Because these orders are parameterization dependent, they should not be interpreted as establishing general superiority of CSRR over ridge regression or other regularization methods.

The numerical studies are designed primarily to verify these limiting properties and to examine the effects of increasing ill conditioning and exact rank deficiency on numerical computation. The accompanying imaginary-component paths are also reported descriptively. No formal inferential or diagnostic interpretation of the imaginary component is proposed in the present work.

The remainder of the paper is organized as follows. Section~\ref{sec:preliminaries} reviews the least-squares and ridge regression formulations used in the subsequent development. Section~\ref{sec:csrr} introduces CSRR, derives its real and imaginary components, and establishes its limiting properties for full-rank and rank-deficient regression problems. Section~\ref{sec:simulation} presents numerical studies illustrating the behavior of CSRR under well-conditioned, increasingly ill-conditioned, and rank-deficient designs. The final section summarizes the main findings and outlines directions for further methodological development.

\section{Preliminaries}
\label{sec:preliminaries}

\subsection{Linear Regression and Least Squares}

Under the linear regression model introduced in equation \eqref{eq:linear-model}, the ordinary least-squares estimator minimizes $\min_{\vecb{\beta}\in\mathbb{R}^{p}}
\left\|
X\vecb{\beta}
-
\vecb{y}
\right\|_2^2.$
%
The corresponding normal equations are $X^{\top}X\vecb{\beta}
=
X^{\top}\vecb{y}.$
%
For notational convenience, define
\begin{equation}
B
=
X^{\top}X,
\qquad
\vecb{c}
=
X^{\top}\vecb{y}.
\label{eq:bc-notation}
\end{equation}
Here, $B\in\mathbb{R}^{p\times p}$ is real symmetric and positive semidefinite, and $\vecb{c}\in\mathbb{R}^{p}$ is the right-hand side of the normal equations. If $X$ has full column rank, then $B$ is positive definite and the least-squares regression coefficient vector is uniquely given by
\begin{equation}
\vecb{\beta}_{\mathrm{LS}}
=
B^{-1}\vecb{c}.
\label{eq:least-squares-solution}
\end{equation}
When the predictors are strongly correlated, $B$ may become severely ill conditioned, causing the least-squares regression coefficient estimates to be sensitive to small perturbations in the data. If $X$ is rank deficient, then $B$ is singular and the least-squares regression coefficient vector is not unique \cite{Bjorck1996,GolubVanLoan2013,Hansen1998}.

\subsection{Ridge Regression}

Ridge regression modifies the least-squares normal equations by adding a positive diagonal shift \cite{HoerlKennard1970,HoerlKennard1970Applications}. For $\mu>0$, the ridge regression coefficient vector satisfies
\begin{equation}
\left(
B+\mu I_p
\right)
\vecb{\beta}_{R,\mu}
=
\vecb{c},
\label{eq:ridge-prelim}
\end{equation}
where $I_p$ is the $p\times p$ identity matrix, $\mu$ is the ridge regularization parameter, and $\vecb{\beta}_{R,\mu}$ denotes the ridge regression coefficient vector. Since $B$ is positive semidefinite, $B+\mu I_p$ is positive definite for every $\mu>0$. Hence ridge regression yields a unique regression coefficient vector even when $B$ is singular.Equivalently, the ridge estimator is the solution of
\begin{equation}
\vecb{\beta}_{R,\mu}
=
\underset{\vecb{\beta}\in\mathbb{R}^{p}}{\arg\min}
\left\{
\left\|
X\vecb{\beta}
-
\vecb{y}
\right\|_2^2
+
\mu
\left\|
\vecb{\beta}
\right\|_2^2
\right\}.
\label{eq:ridge-objective}
\end{equation}
Thus, ridge regression stabilizes estimation by introducing an $\ell_2$ penalty on the regression coefficients, producing a bias--variance tradeoff \cite{HoerlKennard1970,Marquardt1970}. Generalized ridge estimators and methods for selecting the regularization parameter are discussed in \cite{Hemmerle1975,GolubHeathWahba1979}. The next section considers a different perturbation of the normal equations, replacing the real-valued shift $\mu I_p$ by the purely imaginary shift $\ii\mu I_p$.


\section{Complex-Shifted Regularized Regression}
\label{sec:csrr}

\subsection{Formulation of CSRR}

A purely imaginary perturbation has previously been used as a regularization mechanism for function approximation \cite{KuoChenKarageorghis2027}. Motivated by this complex-shift construction, we consider an analogous perturbation of the least-squares normal equations. The resulting \emph{Complex-Shifted Regularized Regression} (CSRR) system is defined by
\begin{equation}
\left(
B+\ii\mu I_p
\right)
\widetilde{\vecb{\beta}}_{\mu}
=
\vecb{c},
\qquad
\mu>0,
\label{eq:csrr-system}
\end{equation}
where $B=X^{\top}X$, $\vecb{c}=X^{\top}\vecb{y}$, $I_p$ is the $p\times p$ identity matrix, $\mu$ is the regularization parameter, $\ii=\sqrt{-1}$, and $\widetilde{\vecb{\beta}}_{\mu}\in\mathbb{C}^{p}$ is the complex-valued CSRR coefficient vector. We write
\begin{equation}
\widetilde{\vecb{\beta}}_{\mu}
=
\vecb{\alpha}_{\mu}
+
\ii\vecb{\gamma}_{\mu},
\qquad
\vecb{\alpha}_{\mu},
\vecb{\gamma}_{\mu}
\in\mathbb{R}^{p},
\label{eq:csrr-components}
\end{equation}
where $\vecb{\alpha}_{\mu}$ and $\vecb{\gamma}_{\mu}$ denote the real and imaginary components, respectively. The real component is used as the regularized regression coefficient vector, while the imaginary component arises from the complex perturbation and is examined descriptively in the numerical studies.

The first basic property of CSRR is that the shifted coefficient matrix is nonsingular for every $\mu>0$, including when $B$ itself is singular.

\begin{proposition}
\label{prop:csrr-nonsingularity}
Let $B=X^{\top}X$ be real symmetric and positive semidefinite. Then
\begin{equation}
B+\ii\mu I_p
\label{eq:csrr-shifted-matrix}
\end{equation}
is nonsingular for every $\mu>0$.
\end{proposition}

\begin{proof}
Because $B$ is real symmetric and positive semidefinite, it admits the spectral decomposition \cite{HornJohnson2012,GolubVanLoan2013}
\begin{equation}
B
=
V\Lambda V^{\top},
\label{eq:csrr-eigendecomposition}
\end{equation}
where $V\in\mathbb{R}^{p\times p}$ is orthogonal and
\begin{equation}
\Lambda
=
\operatorname{diag}
\left(
\lambda_1,\ldots,\lambda_p
\right),
\qquad
\lambda_j\geq0,
\quad
j=1,\ldots,p.
\label{eq:csrr-eigenvalues}
\end{equation}
It follows that
\begin{equation}
B+\ii\mu I_p
=
V
\left(
\Lambda+\ii\mu I_p
\right)
V^{\top}.
\label{eq:csrr-shifted-decomposition}
\end{equation}
Hence, the eigenvalues of $B+\ii\mu I_p$ are
\begin{equation}
\lambda_j+\ii\mu,
\qquad
j=1,\ldots,p.
\label{eq:csrr-shifted-eigenvalues}
\end{equation}
For every $\mu>0$,
\begin{equation}
\left|
\lambda_j+\ii\mu
\right|
=
\sqrt{\lambda_j^2+\mu^2}
>
0.
\label{eq:csrr-eigenvalue-modulus}
\end{equation}
Thus none of the eigenvalues is zero, and therefore
\begin{equation}
\det
\left(
B+\ii\mu I_p
\right)
\neq
0.
\label{eq:csrr-nonsingular}
\end{equation}
Hence, the CSRR system has a unique complex-valued solution for every $\mu>0$, including when $B=X^{\top}X$ is rank deficient.
\end{proof}

\subsection{Real and Imaginary Components}

Substituting \eqref{eq:csrr-components} into \eqref{eq:csrr-system} gives
\begin{equation}
\left(
B+\ii\mu I_p
\right)
\left(
\vecb{\alpha}_{\mu}
+
\ii\vecb{\gamma}_{\mu}
\right)
=
\vecb{c}.
\label{eq:csrr-substitution}
\end{equation}
Separating the real and imaginary parts yields
\begin{equation}
B\vecb{\alpha}_{\mu}
-
\mu\vecb{\gamma}_{\mu}
=
\vecb{c},
\label{eq:csrr-real-equation}
\end{equation}
and
\begin{equation}
B\vecb{\gamma}_{\mu}
+
\mu\vecb{\alpha}_{\mu}
=
\vecb{0}.
\label{eq:csrr-imaginary-equation}
\end{equation}
Thus, the real and imaginary components are coupled through the regularization parameter $\mu$.

Using
\begin{equation}
\left(
B+\ii\mu I_p
\right)
\left(
B-\ii\mu I_p
\right)
=
B^2+\mu^2I_p,
\label{eq:csrr-conjugate-product}
\end{equation}
and noting that $B^2+\mu^2I_p$ is positive definite for every $\mu>0$, we obtain
\begin{equation}
\left(
B+\ii\mu I_p
\right)^{-1}
=
\left(
B-\ii\mu I_p
\right)
\left(
B^2+\mu^2I_p
\right)^{-1}.
\label{eq:csrr-inverse}
\end{equation}
Therefore,
\begin{equation}
\widetilde{\vecb{\beta}}_{\mu}
=
\left(
B-\ii\mu I_p
\right)
\left(
B^2+\mu^2I_p
\right)^{-1}
\vecb{c}.
\label{eq:csrr-explicit-solution}
\end{equation}
Comparing the real and imaginary parts gives
\begin{equation}
\vecb{\alpha}_{\mu}
=
B
\left(
B^2+\mu^2I_p
\right)^{-1}
\vecb{c},
\label{eq:csrr-real-solution}
\end{equation}
and
\begin{equation}
\vecb{\gamma}_{\mu}
=
-\mu
\left(
B^2+\mu^2I_p
\right)^{-1}
\vecb{c}.
\label{eq:csrr-imaginary-solution}
\end{equation}
Hence, both components of the complex-valued CSRR solution admit explicit real-valued representations.

Eliminating $\vecb{\gamma}_{\mu}$ from
\eqref{eq:csrr-real-equation} and
\eqref{eq:csrr-imaginary-equation} gives
\begin{equation}
\left(
B^2+\mu^2I_p
\right)
\vecb{\alpha}_{\mu}
=
B\vecb{c}.
\label{eq:csrr-real-system}
\end{equation}
Equation~\eqref{eq:csrr-real-system} is the normal equation associated with
\begin{equation}
\vecb{\alpha}_{\mu}
=
\underset{\vecb{z}\in\mathbb{R}^{p}}{\arg\min}
\left\{
\left\|
B\vecb{z}
-
\vecb{c}
\right\|_2^2
+
\mu^2
\left\|
\vecb{z}
\right\|_2^2
\right\}.
\label{eq:csrr-real-objective}
\end{equation}
Here, $\vecb{z}\in\mathbb{R}^{p}$ denotes a generic candidate regression coefficient vector. Thus, the real component of CSRR is equivalent to zeroth-order Tikhonov regularization applied to the normal-equation system
\begin{equation}
B\vecb{z}
=
\vecb{c},
\label{eq:normal-equation-system}
\end{equation}
rather than directly to the original least-squares criterion \cite{Tikhonov1963,Gockenbach2016}. This distinction clarifies the relationship between CSRR and classical ridge regression.

\subsection{Limiting Behavior as \texorpdfstring{$\mu\rightarrow0$}{mu to 0}}
\label{sec:csrr-convergence}

\subsubsection{Full-Rank Case}

Suppose first that $X$ has full column rank. Then
$B=X^{\top}X$ is positive definite, and the ordinary least-squares regression coefficient vector is
\begin{equation}
\vecb{\beta}_{\mathrm{LS}}
=
B^{-1}\vecb{c}.
\label{eq:csrr-ls-solution}
\end{equation}

The following result characterizes the limiting behavior of both components of the CSRR solution.

\begin{proposition}
\label{prop:csrr-fullrank-limit}
Let $B=X^{\top}X$ be positive definite, and let $\lambda_{\min}>0$ denote its smallest eigenvalue. Then
\begin{equation}
\left\|
\vecb{\alpha}_{\mu}
-
\vecb{\beta}_{\mathrm{LS}}
\right\|_2
\leq
\frac{\mu^2}
{\lambda_{\min}^2+\mu^2}
\left\|
\vecb{\beta}_{\mathrm{LS}}
\right\|_2,
\label{eq:csrr-fullrank-bound}
\end{equation}
and
\begin{equation}
\left\|
\vecb{\gamma}_{\mu}
\right\|_2
\leq
\frac{\mu}
{\lambda_{\min}}
\left\|
\vecb{\beta}_{\mathrm{LS}}
\right\|_2.
\label{eq:csrr-imaginary-bound}
\end{equation}
Consequently,
\begin{equation}
\vecb{\alpha}_{\mu}
\longrightarrow
\vecb{\beta}_{\mathrm{LS}},
\qquad
\left\|
\vecb{\alpha}_{\mu}
-
\vecb{\beta}_{\mathrm{LS}}
\right\|_2
=
O\left(\mu^2\right),
\qquad
\mu\rightarrow0,
\label{eq:csrr-fullrank-order}
\end{equation}
and
\begin{equation}
\vecb{\gamma}_{\mu}
\longrightarrow
\vecb{0},
\qquad
\left\|
\vecb{\gamma}_{\mu}
\right\|_2
=
O\left(\mu\right),
\qquad
\mu\rightarrow0.
\label{eq:csrr-imaginary-order}
\end{equation}
\end{proposition}

\begin{proof}
Since
$\vecb{c}=B\vecb{\beta}_{\mathrm{LS}}$, equation~\eqref{eq:csrr-real-solution} gives
\begin{equation}
\vecb{\alpha}_{\mu}
=
B^2
\left(
B^2+\mu^2I_p
\right)^{-1}
\vecb{\beta}_{\mathrm{LS}}.
\label{eq:csrr-alpha-ls}
\end{equation}
Using
\begin{equation}
B^2
=
\left(
B^2+\mu^2I_p
\right)
-
\mu^2I_p,
\label{eq:csrr-square-identity}
\end{equation}
we obtain
\begin{equation}
\vecb{\alpha}_{\mu}
-
\vecb{\beta}_{\mathrm{LS}}
=
-\mu^2
\left(
B^2+\mu^2I_p
\right)^{-1}
\vecb{\beta}_{\mathrm{LS}}.
\label{eq:csrr-fullrank-error}
\end{equation}
Because the smallest eigenvalue of $B^2+\mu^2I_p$ is
$\lambda_{\min}^2+\mu^2$,
\begin{equation}
\left\|
\left(
B^2+\mu^2I_p
\right)^{-1}
\right\|_2
=
\frac{1}
{\lambda_{\min}^2+\mu^2},
\label{eq:csrr-fullrank-inverse-norm}
\end{equation}
which gives \eqref{eq:csrr-fullrank-bound}.

Similarly, equation~\eqref{eq:csrr-imaginary-solution} gives
\begin{equation}
\vecb{\gamma}_{\mu}
=
-\mu
\left(
B^2+\mu^2I_p
\right)^{-1}
B
\vecb{\beta}_{\mathrm{LS}}.
\label{eq:csrr-gamma-ls}
\end{equation}
The eigenvalues of
$\left(B^2+\mu^2I_p\right)^{-1}B$
are
$\lambda_j/(\lambda_j^2+\mu^2)$, and for $\lambda_j\geq\lambda_{\min}>0$,
\begin{equation}
\frac{\lambda_j}
{\lambda_j^2+\mu^2}
\leq
\frac{1}{\lambda_j}
\leq
\frac{1}{\lambda_{\min}}.
\label{eq:csrr-gamma-bound-step}
\end{equation}
Therefore, \eqref{eq:csrr-imaginary-bound} follows. The limiting results are immediate.
\end{proof}

\subsubsection{Comparison with Ridge Regression}

For comparison, classical ridge regression satisfies
\begin{equation}
\left(
B+\mu I_p
\right)
\vecb{\beta}_{R,\mu}
=
\vecb{c}.
\label{eq:ridge-comparison-system}
\end{equation}
For a full-rank problem, substituting
$\vecb{c}=B\vecb{\beta}_{\mathrm{LS}}$ yields
\begin{equation}
\vecb{\beta}_{R,\mu}
-
\vecb{\beta}_{\mathrm{LS}}
=
-\mu
\left(
B+\mu I_p
\right)^{-1}
\vecb{\beta}_{\mathrm{LS}}.
\label{eq:ridge-error}
\end{equation}
Consequently,
\begin{equation}
\left\|
\vecb{\beta}_{R,\mu}
-
\vecb{\beta}_{\mathrm{LS}}
\right\|_2
\leq
\frac{\mu}
{\lambda_{\min}+\mu}
\left\|
\vecb{\beta}_{\mathrm{LS}}
\right\|_2,
\label{eq:ridge-error-bound}
\end{equation}
and therefore
\begin{equation}
\left\|
\vecb{\beta}_{R,\mu}
-
\vecb{\beta}_{\mathrm{LS}}
\right\|_2
=
O\left(\mu\right),
\qquad
\mu\rightarrow0.
\label{eq:ridge-error-order}
\end{equation}

Thus, for a fixed full-rank regression problem and under the same numerical parameterization, the CSRR real component has an $O(\mu^2)$ perturbation from the least-squares regression coefficient vector, whereas classical ridge regression with shift $\mu I_p$ has an $O(\mu)$ perturbation.

This comparison concerns the local dependence of the two estimators on their stated regularization parameters. The orders $O(\mu^2)$ and $O(\mu)$ are parameterization dependent and are not invariant to reparameterization. Accordingly, this result should not be interpreted as establishing general superiority of CSRR over ridge regression or other regularization methods.

\subsubsection{Rank-Deficient Case}

Suppose now that $X$ is rank deficient, so that
$B=X^{\top}X$ is singular and the least-squares regression coefficient vector is not unique. A natural reference is the Moore--Penrose minimum-norm least-squares solution \cite{Penrose1955,Bjorck1996},
\begin{equation}
\vecb{\beta}_{\mathrm{LS}}^{+}
=
B^{+}\vecb{c},
\label{eq:csrr-minimum-norm}
\end{equation}
where $B^{+}$ denotes the Moore--Penrose generalized inverse of $B$ and $\vecb{\beta}_{\mathrm{LS}}^{+}$ denotes the minimum-norm least-squares regression coefficient vector.

\begin{proposition}
\label{prop:csrr-minimum-norm}
Let $B=X^{\top}X$ be real symmetric and positive semidefinite, possibly singular. Then
\begin{equation}
\lim_{\mu\rightarrow0}
\vecb{\alpha}_{\mu}
=
\vecb{\beta}_{\mathrm{LS}}^{+},
\label{eq:csrr-minimum-norm-limit}
\end{equation}
and
\begin{equation}
\lim_{\mu\rightarrow0}
\vecb{\gamma}_{\mu}
=
\vecb{0}.
\label{eq:csrr-rank-deficient-imaginary-limit}
\end{equation}
\end{proposition}

\begin{proof}
Let
\begin{equation}
B
=
V\Lambda V^{\top}
\label{eq:csrr-rank-deficient-decomposition}
\end{equation}
be the spectral decomposition of $B$. Since
$\vecb{c}=X^{\top}\vecb{y}$ and
$\mathcal{R}(X^{\top}X)=\mathcal{R}(X^{\top})$,
$
\vecb{c}
\in
\mathcal{R}(B),
\label{eq:csrr-range-condition}$
where $\mathcal{R}(B)$ denotes the range of $B$. Hence, the coordinates of $V^{\top}\vecb{c}$ corresponding to zero eigenvalues of $B$ are zero. From \eqref{eq:csrr-real-solution},
\begin{equation}
\vecb{\alpha}_{\mu}
=
V
\operatorname{diag}
\left(
\frac{\lambda_1}{\lambda_1^2+\mu^2},
\ldots,
\frac{\lambda_p}{\lambda_p^2+\mu^2}
\right)
V^{\top}\vecb{c}.
\label{eq:csrr-rank-deficient-real}
\end{equation}
For each $\lambda_j>0$,
\begin{equation}
\frac{\lambda_j}
{\lambda_j^2+\mu^2}
\longrightarrow
\frac{1}{\lambda_j},
\qquad
\mu\rightarrow0.
\label{eq:csrr-positive-eigenvalue-limit}
\end{equation}
The coordinates corresponding to $\lambda_j=0$ contribute zero because
$\vecb{c}\in\mathcal{R}(B)$. Hence,
\begin{equation}
\vecb{\alpha}_{\mu}
\longrightarrow
B^{+}\vecb{c}
=
\vecb{\beta}_{\mathrm{LS}}^{+}.
\label{eq:csrr-rank-deficient-real-limit}
\end{equation}

Likewise, from \eqref{eq:csrr-imaginary-solution},
\begin{equation}
\vecb{\gamma}_{\mu}
=
-
V
\operatorname{diag}
\left(
\frac{\mu}{\lambda_1^2+\mu^2},
\ldots,
\frac{\mu}{\lambda_p^2+\mu^2}
\right)
V^{\top}\vecb{c}.
\label{eq:csrr-rank-deficient-imaginary}
\end{equation}
For every $\lambda_j>0$, the corresponding multiplier converges to zero as $\mu\rightarrow0$, while the coordinates associated with $\lambda_j=0$ again contribute zero. Therefore,
$\vecb{\gamma}_{\mu}\rightarrow\vecb{0}$.
\end{proof}

Proposition~\ref{prop:csrr-minimum-norm} establishes that the real component of CSRR converges to the Moore--Penrose minimum-norm least-squares solution when the design matrix is rank deficient. Thus, CSRR provides a uniquely defined regularized regression coefficient vector for every $\mu>0$ and a well-defined limiting solution as $\mu\rightarrow0$.

The preceding results establish the fundamental algebraic and limiting properties required for the numerical studies that follow. The imaginary component is examined descriptively as part of the complex-valued solution; a formal statistical interpretation of this component is beyond the scope of the present paper.

\section{Numerical Studies}
\label{sec:simulation}

This section examines the numerical behavior of Complex-Shifted Regularized Regression (CSRR) under well-conditioned, increasingly ill-conditioned, and exactly rank-deficient regression designs. The numerical examples are intended to illustrate the theoretical results established in Section~\ref{sec:csrr}, rather than to provide a comprehensive comparison of predictive performance among regularization methods. 

In particular, we examine convergence of the CSRR real component toward the ordinary least-squares solution in full-rank problems and toward the Moore--Penrose minimum-norm least-squares solution in rank-deficient problems. The imaginary component is examined descriptively as part of the complex-valued solution.

For full-rank problems, coefficient error is measured relative to the ordinary least-squares solution as
\begin{equation}
E_{\beta}^{\mathrm{CSRR}}(\mu)
=
\left\|
\vecb{\alpha}_{\mu}
-
\vecb{\beta}_{\mathrm{LS}}
\right\|_2
\label{eq:csrr-coefficient-error}
\end{equation}
for CSRR and
\begin{equation}
E_{\beta}^{\mathrm{Ridge}}(\mu)
=
\left\|
\vecb{\beta}_{R,\mu}
-
\vecb{\beta}_{\mathrm{LS}}
\right\|_2
\label{eq:ridge-coefficient-error}
\end{equation}
for ridge regression. For rank-deficient designs, the Moore--Penrose minimum-norm least-squares solution is used as the reference. The error measures therefore quantify numerical convergence toward the appropriate least-squares solution and are not measures of out-of-sample prediction performance.

All numerical calculations were performed in R using double-precision arithmetic (R Core Team, 2025)~\cite{RCoreTeam2025}. The polynomial-regression example used $n=30$ equally spaced points on $[-1,1]$ and 300 logarithmically spaced values of $\mu$ from $10^{0}$ to $10^{-15}$. The multiple-regression example used $n=100$ observations and 400 logarithmically spaced values of $\mu$ over the same interval. A fixed random seed of 20261110 was used for the stochastic multiple-regression example. For the exactly rank-deficient design, the Moore--Penrose reference solution and the CSRR and ridge regularization paths were evaluated using the singular value decomposition of the centered predictor matrix \cite{GolubVanLoan2013}. Singular values were treated as numerically nonzero when they exceeded $\tau
=
\max(n,p)\,
d_{\max}\,
\epsilon_{\mathrm{mach}},
\label{eq:svd-tolerance}$
where $d_{\max}$ is the largest singular value and $\epsilon_{\mathrm{mach}}$ is the double-precision machine epsilon. This implementation avoids direct inversion of numerically singular normal-equation matrices.

\subsection{Polynomial Regression with Increasing Ill Conditioning}
\label{sec:polynomial-example}

We first examine the numerical behavior of CSRR using polynomial approximation of
\begin{equation}
f(x)
=
\exp(x),
\qquad
x\in[-1,1].
\label{eq:polynomial-function}
\end{equation}
We use $n=30$ equally spaced points on $[-1,1]$ and define
\begin{equation}
y_i
=
\exp(x_i),
\qquad
i=1,\ldots,n.
\label{eq:polynomial-response}
\end{equation}
For a polynomial of degree $k$, the predictor matrix contains the powers
$x_i,x_i^2,\ldots,x_i^k$. Thus,
\begin{equation}
X
=
\begin{bmatrix}
x_1 & x_1^2 & \cdots & x_1^k\\
x_2 & x_2^2 & \cdots & x_2^k\\
\vdots & \vdots & \ddots & \vdots\\
x_n & x_n^2 & \cdots & x_n^k
\end{bmatrix}.
\label{eq:polynomial-design}
\end{equation}
The columns of $X$ and the response vector are centered before fitting, consistent with the formulation in Sections~\ref{sec:preliminaries} and~\ref{sec:csrr}. The intercept is therefore treated separately and is not regularized.

The purpose of this example is twofold. First, the well-conditioned case is used to verify numerically the local perturbation orders established in Section~\ref{sec:csrr-convergence}. Second, higher-degree polynomial models are used to examine how increasing ill conditioning affects the attainable numerical accuracy of CSRR and ridge regression.

For full-rank designs, coefficient error is measured relative to the ordinary least-squares solution by
\begin{equation}
E_{\beta}^{\mathrm{CSRR}}(\mu)
=
\left\|
\vecb{\alpha}_{\mu}
-
\vecb{\beta}_{\mathrm{LS}}
\right\|_2
\label{eq:polynomial-csrr-error}
\end{equation}
for CSRR and
\begin{equation}
E_{\beta}^{\mathrm{Ridge}}(\mu)
=
\left\|
\vecb{\beta}_{R,\mu}
-
\vecb{\beta}_{\mathrm{LS}}
\right\|_2
\label{eq:polynomial-ridge-error}
\end{equation}
for ridge regression. Under the common numerical parameterization used here, Section~\ref{sec:csrr-convergence} establishes local perturbation orders of $O(\mu^2)$ for the CSRR real component and $O(\mu)$ for ridge regression as $\mu\rightarrow0$.

\subsubsection{Well-Conditioned Case}

We first consider the first-degree polynomial model, $k=1$. For this design,
\begin{equation}
\lambda_{\min}(B)
=
10.69,
\qquad
\kappa_2(B)
=
1,
\label{eq:k1-conditioning}
\end{equation}
where $B=X^{\top}X$ and $\kappa_2(B)$ denotes the spectral condition number.

\begin{figure}[h!]
\centering
\includegraphics[width=0.55\textwidth]
{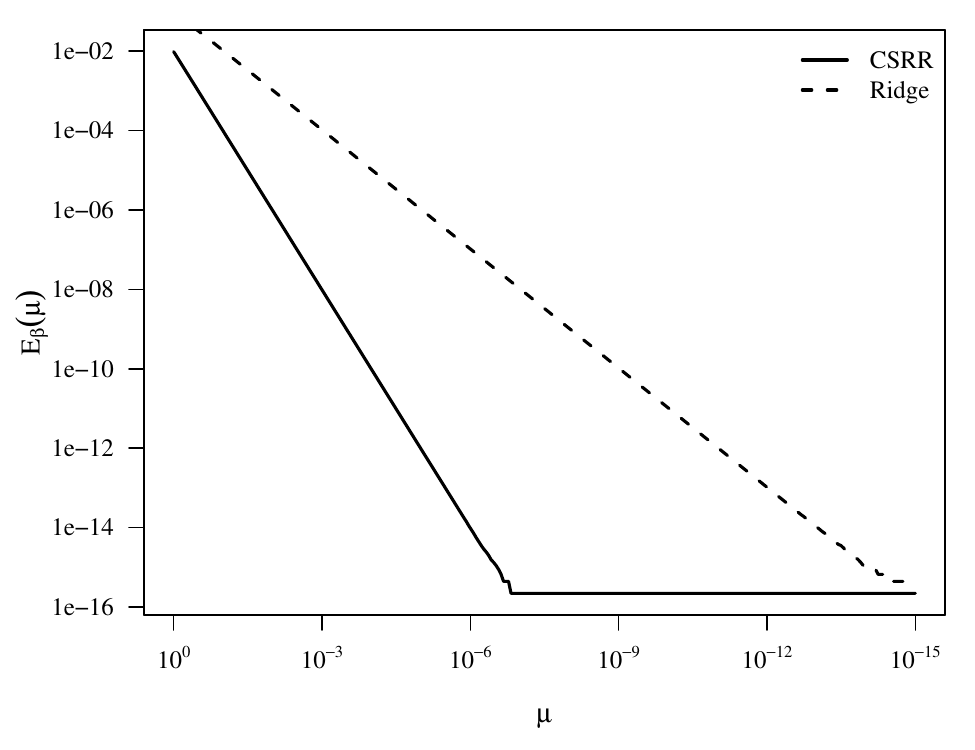}
\caption{Coefficient error as a function of the regularization
parameter $\mu$ for the first-degree polynomial example
($k=1$). The solid and dashed curves represent CSRR and
ridge regression, respectively. The regularization parameter
decreases from left to right. Both methods approach the
least-squares solution as $\mu$ decreases, with the computed
errors eventually reaching finite-precision levels.}
\label{fig:polynomial-k1-error}
\end{figure}

Figure~\ref{fig:polynomial-k1-error} shows the coefficient errors for CSRR and ridge regression over decreasing values of $\mu$. Both regularized solutions approach the ordinary least-squares solution. To quantify the observed local behavior, a linear regression of
$\log_{10}E_{\beta}(\mu)$ on $\log_{10}\mu$ was fitted over the numerically stable small-$\mu$ region. The estimated regression coefficient was $1.99997$ for CSRR and $0.99999$ for ridge regression, with $R^2>0.99999$ for both fits. These numerical regression coefficients agree closely with the theoretical orders of two and one, respectively.

As $\mu$ decreases further, the CSRR error reaches approximately $2.22\times10^{-16}$, after which finite-precision effects dominate and further reduction in $\mu$ does not materially reduce the computed error.

For the first-degree polynomial example, the magnitude of the imaginary component decreased smoothly toward zero as \(\mu\) decreased, consistent with \eqref{eq:csrr-imaginary-order}.


\subsubsection{Increasingly Ill-Conditioned Cases}

We next consider polynomial degrees $k=8$ and $k=11$. Increasing the polynomial degree produces increasingly dependent columns in the polynomial design matrix and substantially worsens the conditioning of the normal equations. Table~\ref{tab:polynomial-conditioning} summarizes the resulting numerical characteristics.

\begin{table}[htbp]
\centering
\caption{Conditioning and minimum observed coefficient errors for the polynomial regression examples.}
\label{tab:polynomial-conditioning}
\begin{tabular}{ccccc}
\hline
$k$
&
$\lambda_{\min}(B)$
&
$\kappa_2(B)$
&
Minimum CSRR error
&
Minimum ridge error
\\
\hline
1
&
$1.069\times10^{1}$
&
$1.00$
&
$2.22\times10^{-16}$
&
$4.44\times10^{-16}$
\\
8
&
$1.485\times10^{-4}$
&
$1.405\times10^{5}$
&
$2.13\times10^{-12}$
&
$1.27\times10^{-12}$
\\
11
&
$7.736\times10^{-7}$
&
$3.210\times10^{7}$
&
$2.29\times10^{-9}$
&
$7.94\times10^{-10}$
\\
\hline
\end{tabular}
\end{table}

For $k=8$, the smallest eigenvalue decreases to approximately
$1.48\times10^{-4}$ and the condition number increases to
$1.40\times10^{5}$. For $k=11$, the smallest eigenvalue decreases further to
$7.74\times10^{-7}$, while the condition number increases to approximately
$3.21\times10^{7}$. Thus, the three polynomial examples provide a progression from a well-conditioned problem to increasingly severe ill-conditioning.

\begin{figure}[h!]
\centering
\includegraphics[width=0.92\textwidth]
{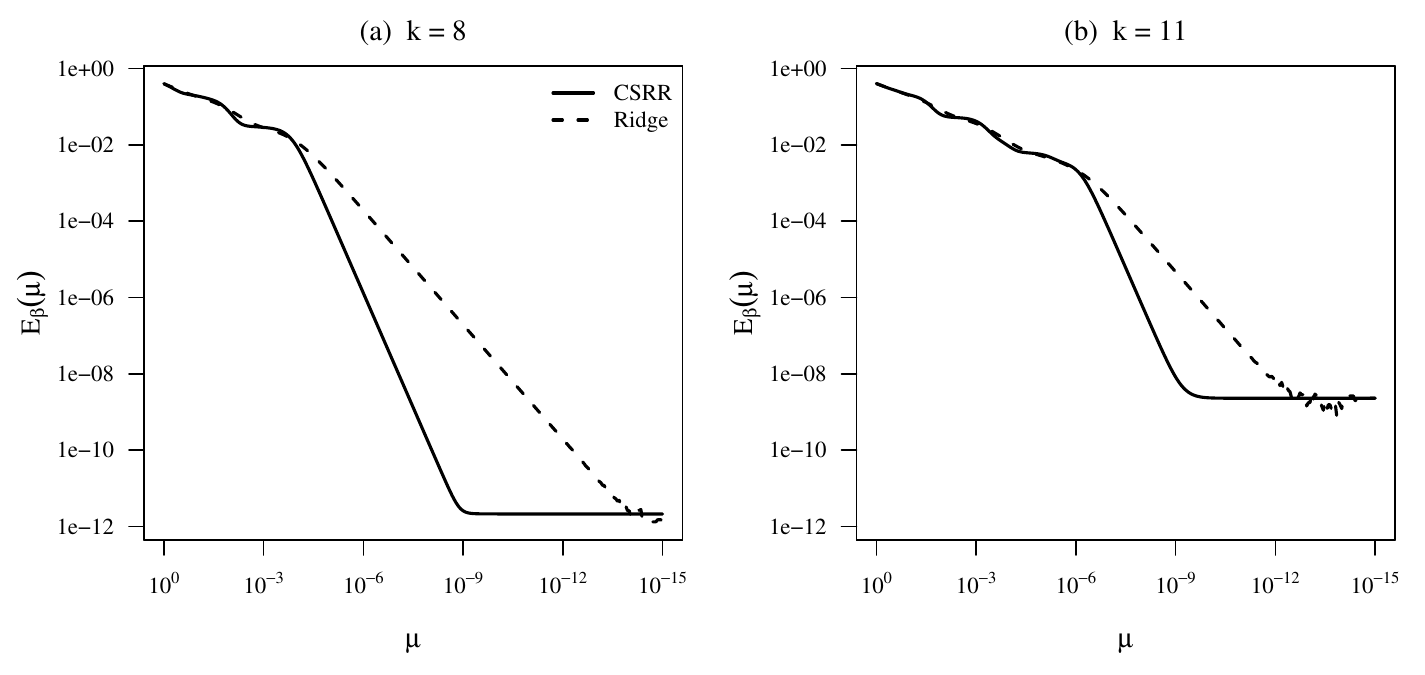}
\caption{Coefficient error as a function of the regularization parameter $\mu$ for increasingly ill-conditioned polynomial regression problems: (a) $k=8$ and (b) $k=11$. Solid curves represent the real component of CSRR and dashed curves represent ridge regression. The regularization parameter decreases from left to right. As the polynomial degree increases and the normal-equation matrix becomes more ill conditioned, finite-precision effects occur at progressively larger coefficient-error levels.}
\label{fig:polynomial-high-degree}
\end{figure}

Figure~\ref{fig:polynomial-high-degree} shows the coefficient-error trajectories for the increasingly ill-conditioned polynomial designs. For $k=8$, the smallest eigenvalue of $B=X^{\top}X$ was $1.48\times10^{-4}$ and the spectral condition number was $1.40\times10^{5}$. For $k=11$, these values were $7.74\times10^{-7}$ and $3.21\times10^{7}$, respectively. Thus, increasing the polynomial degree produced a substantial deterioration in the conditioning of the normal equations.

For both designs, the CSRR and ridge solutions approach the ordinary least-squares solution as $\mu$ decreases. However, the attainable numerical accuracy deteriorates as the conditioning worsens. The minimum observed CSRR coefficient error increased from $2.13\times10^{-12}$ for $k=8$ to $2.29\times10^{-9}$ for $k=11$. The corresponding minimum ridge errors were $1.27\times10^{-12}$ and $7.94\times10^{-10}$, respectively. At sufficiently small values of $\mu$, the error trajectories flatten or fluctuate because finite-precision effects dominate the computation.

Unlike the well-conditioned $k=1$ example, these higher-degree cases are not used to estimate empirical convergence orders. Severe ill conditioning narrows the numerically observable asymptotic region and causes finite-precision effects to appear before a stable log--log convergence regime can be estimated over a broad range of $\mu$.

The polynomial examples therefore illustrate two distinct properties of the proposed method. In the well-conditioned setting, the numerical behavior closely reproduces the local perturbation orders derived in Section~\ref{sec:csrr-convergence}. As the design becomes increasingly ill conditioned, the attainable numerical accuracy deteriorates and finite-precision effects become visible at larger error levels. These findings distinguish the theoretical limiting behavior as $\mu\rightarrow0$ from the numerical accuracy that can be achieved in finite-precision computation.

\subsection{Multiple Regression with Independent and Exactly Dependent Predictors}
\label{sec:multiple-regression-example}

We next examine CSRR in a multivariable regression setting under two predictor structures. The first case uses independent predictors and provides a well-conditioned, full-rank baseline. The second introduces an exact linear dependence, producing a rank-deficient design for which the ordinary least-squares regression coefficient vector is not unique. The primary objective is to illustrate the limiting results established in Section~\ref{sec:csrr-convergence}. The imaginary components are examined descriptively as part of the complex-valued CSRR solution.

\subsubsection{Independent Predictors}

We first generate four mutually independent predictors, $
x_{j,i}
\overset{\mathrm{ind}}{\sim}
N(0,1),
\quad
j=1,\ldots,4,
\quad
i=1,\ldots,n,
\label{eq:independent-predictors}
$
with $n=100$. Responses are generated according to
\begin{equation}
y_i
=
1
+
3x_{1,i}
+
2x_{2,i}
+
1.5x_{3,i}
+
x_{4,i}
+
\varepsilon_i,
\qquad
i=1,\ldots,n,
\label{eq:independent-response}
\end{equation}
where $
\varepsilon_i
\overset{\mathrm{ind}}{\sim}
N(0,0.1^2).
\label{eq:simulation-error}
$
The corresponding regression coefficient vector is
\begin{equation}
\vecb{\beta}^{*}
=
\begin{bmatrix}
3 & 2 & 1.5 & 1
\end{bmatrix}^{\top}.
\label{eq:true-beta-independent}
\end{equation}
As throughout the paper, the predictors and response are centered before regularization, and the intercept is treated separately. Because the resulting design is full rank, the ordinary least-squares solution is used as the reference coefficient vector. For the realized design,
\begin{equation}
\operatorname{rank}(X)
=
4,
\qquad
\kappa_2(B)
=
1.745,
\label{eq:independent-conditioning}
\end{equation}
where $B=X^{\top}X$. Thus, this case represents a well-conditioned multivariable regression problem. At $\mu\approx10^{-6}$, the CSRR coefficient error was $9.42\times10^{-16}$, compared with $3.80\times10^{-8}$ for ridge regression. This behavior is consistent with the different local perturbation orders derived in Section~\ref{sec:csrr-convergence} under the common numerical parameterization used in this study. For smaller values of $\mu$, the CSRR error reaches the finite-precision regime.

\subsubsection{Exact Collinearity}

We next retain the four predictors from the preceding case and introduce a fifth predictor defined by
$
x_5
=
x_1+x_2.
\label{eq:exact-collinearity}
$
The response continues to be generated from equation \eqref{eq:independent-response}. Thus, one valid representation of the regression coefficient vector is
\begin{equation}
\vecb{\beta}^{*}
=
\begin{bmatrix}
3 & 2 & 1.5 & 1 & 0
\end{bmatrix}^{\top}.
\label{eq:true-beta-collinear}
\end{equation}
However, because $
x_1+x_2-x_5
=
0,
\label{eq:dependence-relation}
$
the five regression coefficients are not uniquely identifiable. Different coefficient vectors can produce the same fitted response, and $B=X^{\top}X$ is singular. Accordingly, the Moore--Penrose minimum-norm least-squares solution is used as the reference \cite{Moore1920,Penrose1955,Bjorck1996},
\begin{equation}
\vecb{\beta}_{\mathrm{LS}}^{+}
=
X^{+}\vecb{y}
=
B^{+}\vecb{c}.
\label{eq:simulation-minimum-norm}
\end{equation}

Because the resulting design is full rank, coefficient errors are calculated relative to the ordinary least-squares solution using \eqref{eq:csrr-coefficient-error} and \eqref{eq:ridge-coefficient-error}.


The five-column predictor matrix had rank four and nullity one. Consequently, the ordinary spectral condition number of $B$ is infinite. Among the four identifiable eigendirections, the ratio of the largest to smallest positive eigenvalue was $4.224$, indicating that the numerical difficulty in this example is attributable to exact rank deficiency rather than severe ill conditioning within the identifiable subspace.

For the realized data, the Moore--Penrose minimum-norm solution was
\begin{equation}
\vecb{\beta}_{\mathrm{LS}}^{+}
=
\begin{bmatrix}
1.3271 &
0.3405 &
1.4951 &
1.0140 &
1.6677
\end{bmatrix}^{\top}.
\label{eq:minimum-norm-numerical}
\end{equation}
Although this vector differs from the generating representation in
\eqref{eq:true-beta-collinear}, the identifiable combined effects satisfy
\begin{equation}
\beta_1^{+}+\beta_5^{+}
=
2.995,
\qquad
\beta_2^{+}+\beta_5^{+}
=
2.008,
\label{eq:minimum-norm-combined-effects}
\end{equation}
which are close to the generating coefficients $3$ and $2$, respectively.

\begin{figure}[h!]
    \centering
    \includegraphics[width=0.85\linewidth]{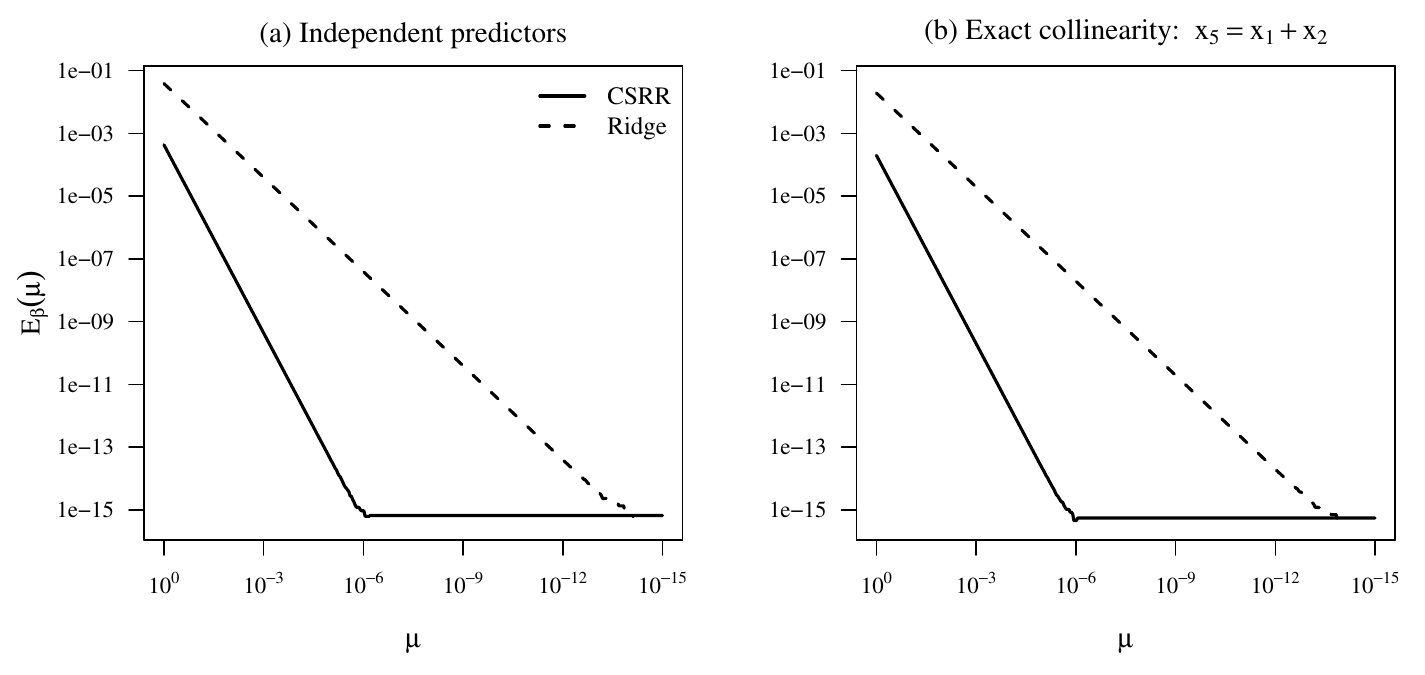}
    \caption{Coefficient error as a function of the regularization parameter $\mu$ for the multiple-regression examples: (a) four independent predictors and (b) exact collinearity with $x_5=x_1+x_2$. Solid curves represent the real component of CSRR and dashed curves represent ridge regression. In panel (a), errors are measured relative to the ordinary least-squares solution; in panel (b), they are measured relative to the Moore--Penrose minimum-norm least-squares solution. The regularization parameter decreases from left to right.}
    \label{fig:multiple-regression-error}
\end{figure}

Figure~\ref{fig:multiple-regression-error} compares the coefficient-error trajectories for the two predictor structures. In the independent-predictor case, errors are measured relative to the ordinary least-squares solution, whereas in the exact-collinearity case they are measured relative to the Moore--Penrose minimum-norm solution. For the rank-deficient design, the CSRR coefficient error at $\mu\approx10^{-6}$ was $4.58\times10^{-16}$, compared with $1.89\times10^{-8}$ for ridge regression. Thus, under the common numerical parameterization used here, the CSRR real component reaches the finite-precision regime at a larger value of $\mu$ than ridge regression. This parameterization-dependent behavior should not be interpreted as establishing general superiority of CSRR over ridge regression.

The rank-deficient case provides a direct numerical illustration of Proposition~\ref{prop:csrr-minimum-norm}: despite the nonuniqueness of the ordinary least-squares regression coefficient vector, the real component of CSRR approaches the uniquely defined Moore--Penrose minimum-norm solution as $\mu\rightarrow0$.

\begin{figure}[h!]
    \centering
    \includegraphics[width=0.85\linewidth]{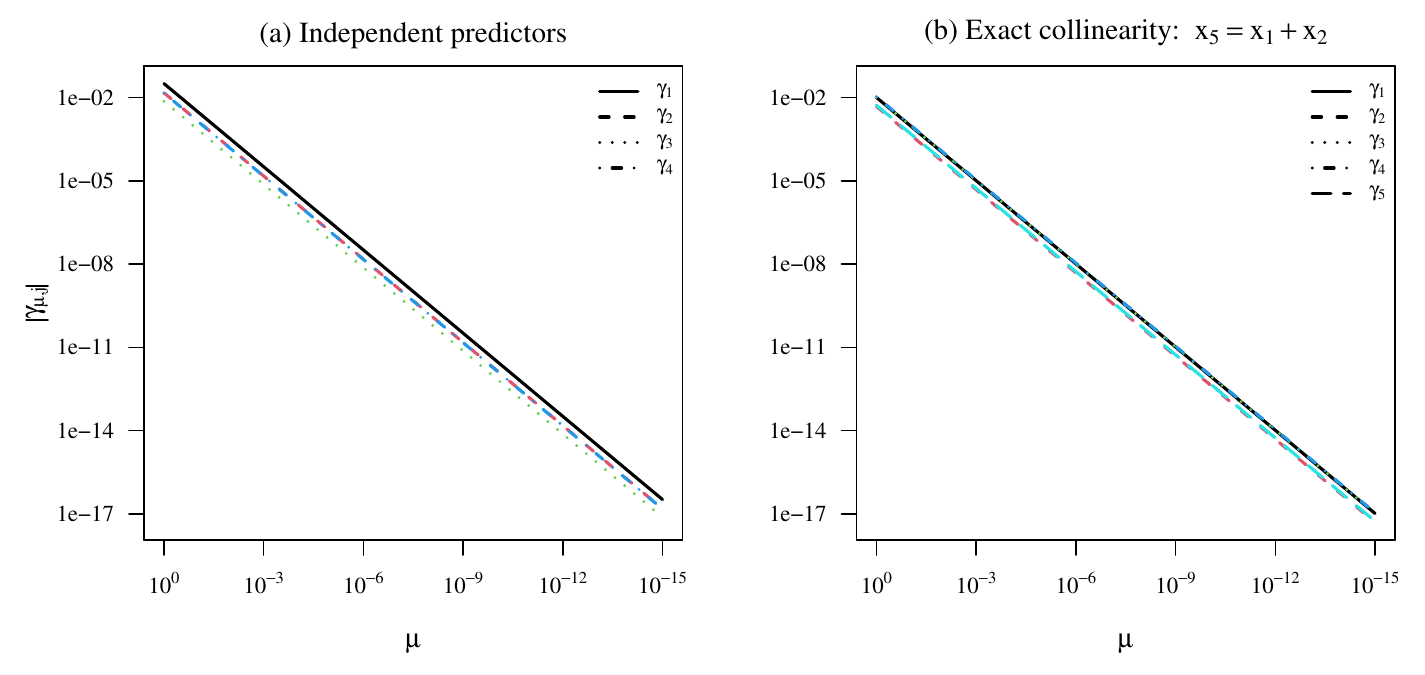}
    \caption{Magnitudes of the imaginary components of the CSRR solution as functions of the regularization parameter $\mu$ for the multiple-regression examples: (a) four independent predictors and (b) exact collinearity with $x_5=x_1+x_2$. The regularization parameter decreases from left to right. In both cases, the imaginary components decrease smoothly toward zero as $\mu$ decreases, consistent with the limiting properties of CSRR.}
    \label{fig:multiple-regression-imaginary}
\end{figure}

Figure~\ref{fig:multiple-regression-imaginary} shows the magnitudes of the imaginary components for the two predictor structures. In both cases, the components decrease smoothly toward zero as $\mu$ decreases. For the full-rank design, this behavior is consistent with \eqref{eq:csrr-imaginary-order}; for the rank-deficient design, it is consistent with Proposition~\ref{prop:csrr-minimum-norm}.

In the rank-deficient realization, the imaginary-component paths do not exhibit a clear separation between predictors involved in the exact dependence relation and the remaining predictors. Accordingly, this example does not support interpreting the imaginary component as a formal diagnostic for collinearity. The imaginary paths are reported only as a descriptive feature of the complex-valued CSRR solution.

The numerical studies illustrate the principal theoretical properties of CSRR. In full-rank problems, the real component approaches the ordinary least-squares solution with the local perturbation behavior derived in Section~\ref{sec:csrr-convergence}. Increasing ill-conditioning limits the attainable accuracy in finite-precision computation, while exact rank deficiency changes the appropriate limiting target from ordinary least squares to the Moore--Penrose minimum-norm solution. The imaginary components converge toward zero in both settings, consistent with the theoretical results.

\section{Conclusions}
\label{sec:conclusions}

This paper introduced Complex-Shifted Regularized Regression (CSRR), a regularization framework for ill-conditioned and rank-deficient linear regression based on a purely imaginary perturbation of the least-squares normal equations. The resulting system is nonsingular for every $\mu>0$, including when $X^{\top}X$ is singular. Explicit expressions were derived for the real and imaginary components of the complex-valued solution, and the real component was shown to admit a Tikhonov-type regularization interpretation.

For full-rank designs, the real component converges to the ordinary least-squares regression coefficient vector as $\mu\rightarrow0$, with local perturbation of order $O(\mu^2)$ under the stated parameterization, compared with $O(\mu)$ for classical ridge regression with shift $\mu I_p$. For rank-deficient designs, the real component converges to the Moore--Penrose minimum-norm least-squares solution. These results establish a well-defined limiting behavior for CSRR in both nonsingular and singular regression problems. The comparison with ridge regression is parameterization dependent and should not be interpreted as establishing general superiority of CSRR.

The numerical studies supported these theoretical properties and illustrated the effects of increasing ill conditioning, exact rank deficiency, and finite-precision computation. The imaginary components decreased toward zero as predicted by the theory and are treated here only as a descriptive feature of the complex-valued solution. Further work may investigate tuning-parameter selection, statistical risk properties, and broader inferential interpretations of CSRR.



\bibliographystyle{unsrt}  
\bibliography{references}

\end{document}